\documentclass{elsart}

 \usepackage{graphicx}
\usepackage{amssymb}

\begin{document}

\begin{frontmatter}



\title{Mobility of O$_{2}^{-}$ ions in supercritical Ar: Experiment and Molecular Dynamics Simulations}


\author{A. F. Borghesani}

\address{Department of Physics and CNISM-Unit, University of Padua, Padua, Italy}
\ead{borghesani@padova.infm.it}

\begin{abstract}
A new analysis and new Molecular Dynamics (MD) simulations of the measurements of the mobility $\mu_{i}$ of O$_{2}^{-} $ ions in dense supercritical Ar gas are reported. $\mu_{i}$ shows a marked dependence on the distance from the critical temperature $T_{c}.$ A mobility defect appears as a function of the gas density and its maximum value occurs below the critical density. The locus of points of maximum mobility defect in the $P-T$ plane appears on the extrapolation of the coexistence curve into the single-phase region. MD simulations quantitatively reproduce the mobility defect near $T_{c}.$
\end{abstract}

\end{frontmatter}

\section{Introduction}
\label{sec:intro}
Negative ions in supercritical gases are an interesting subject for several reasons. 
The formation of anions as a consequence of low-energy electron impact to a molecule has important consequences in the chemistry of atmosphere, of discharges, and in electro-- and biochemistry. 
Electron attachment and detachment processes, leading to the formation and decay of anions are strongly influenced by the environment \cite{illenberger1994b}. 
Upon creation and stabilization, they 
interact with the environment leading to states that cannot be adiabatically obtained by a simple addition of the ion to the surrounding system \cite{hernandez1991,khrapak1999b}.

The investigation of the transport properties of ions in a dense supercritical gas gives important pieces of information about their structure and interaction with the host \cite{borg1993} and bridge the gap between 
 dilute gases \cite{mason1988} and liquids \cite{IS2005}. 

Anions are 
difficult to produce 
even in a dense environment, with the noticeable exception of oxygen and SF$_{6}$ that are very electronegative. In particular, O$_{2}$ is an ubiquitous species, present as impurity even in the best purified rare gas. Resonant electron attachment 
\cite{bloch1935}, followed by stabilization induced by three--body collisions 
 \cite{borghesani1997att}, produces stable O$_{2}^{-}$ ions that can be easily drifted through the gas under the action of an externally applied 
 electric field. 

The O$_{2}^{-}$ ion couples  with the gas by means of electrostriction \cite{atkins1959}. Its strong field polarizes the surrounding atoms 
which are attracted toward the ion that becomes solvated in a cluster. 
 Electrostriction thus induces rapidly varying density and viscosity profiles around the ion \cite{borg1993,ostermeier1972,khrapak1995} that affect 
 the ion mobility $\mu_{i}.$ In these conditions, the ion transport is actually determined by the hydrodynamic interaction of the large solvation structure surrounding the ion with the gas and is quite insensitive to the ionic species.

Whereas 
 $\mu_{i}$ in He at $T=77.2\,$ K \cite{bartels1975} does not show any particular features as a function of the gas density $N$ up to moderate values, it shows a smooth change of behavior when $N$ is varied across the critical density $N_{c}
 $ in Ne gas $0.6\,$K above the critical temperature $T_{c}
 $ \cite{borg1993}. The dependence of $\mu_{i}$ on $N$ for $N> N_{c}$ has been described in terms of the Stokes' hydrodynamic formula relating $\mu_{i}$ to ion radius $R$ and fluid viscosity $\eta:$ 
$ \mu_{i}= 
{e}/{6\pi\eta R},$
in which $\eta$ is evaluated for the density of the maximum of the density profile induced by electrostriction \cite{khrapak1995b}.

Further measurements in near critical Ar gas \cite{borghesani1997} have confirmed that the coupling of the ion motion with the critical fluctuations via electrostriction is responsible for the change of behavior of $\mu_{i}$ near the critical point. The behavior close to $T_{c}$ has been explained by modifying the Stokes's formula in order to account for the fact that the large gas compressibility forces a layer of correlated fluid to stick around the ion. In order to test this interpretation at microscopic level, Molecular Dynamics (MD) simulations are performed.
\section{Experimental results and MD simulations}\label{sec:expMD}
\begin{figure}[t!]
\begin{center}
\includegraphics*[width=7.5cm,angle=-90]{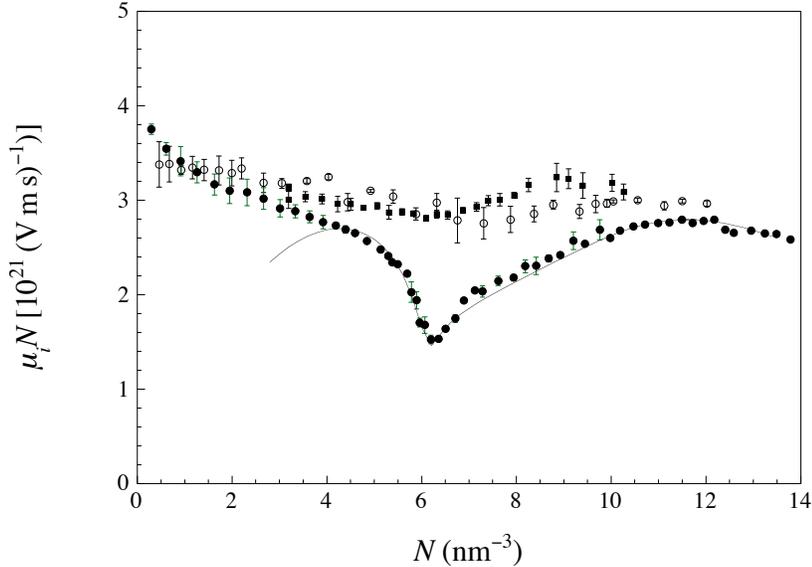}
\end{center}
\caption{$\mu_{i}N $ vs $N.$ Open circles: $T=157\,$K. Squares: $T=154\,$K. Closed circles: $T=151.5\,$K. Solid line: model prediction.\label{fig:misure}}
\end{figure}

The experimental results of the measurements of the O$_{2}^{-}$ ion density-normalized mobility $\mu_{i}N$ in dense supercritical Ar for $T=157,$ $154,$ and $151.5\,$K are shown in fig. \ref{fig:misure}. Details on the experiment and on the previous analysis are found in literature \cite{borghesani1997}.
For a density $N_{m}<N_{c}=8.08\,$nm$^{-3}$ $\mu_{i}$ shows a defect whose strength increases when $T\to T_{c}=150.86\,$K. The solid line through the data on the isotherm closest to $T_{c}$ is obtained by using the Stokes' formula with an ion radius $R$ that is enhanced with respect to the radius in absence of electrostriction and criticality by an amount that depends on the local value of the fluid correlation lenght $\xi$ \cite{borghesani1997}. The good agreement confirms the picture of a correlated fluid layer dragged by the ion because of electrostriction.
For $N<4\,$nm$^{-3}$ the hydrodynamic description fails to reproduce the data because the ion transport leaves the hydrodynamic regime to enter the kinetic one.

\begin{figure}[t!]
\begin{center}
\includegraphics*[width=7.5cm,angle=-90]{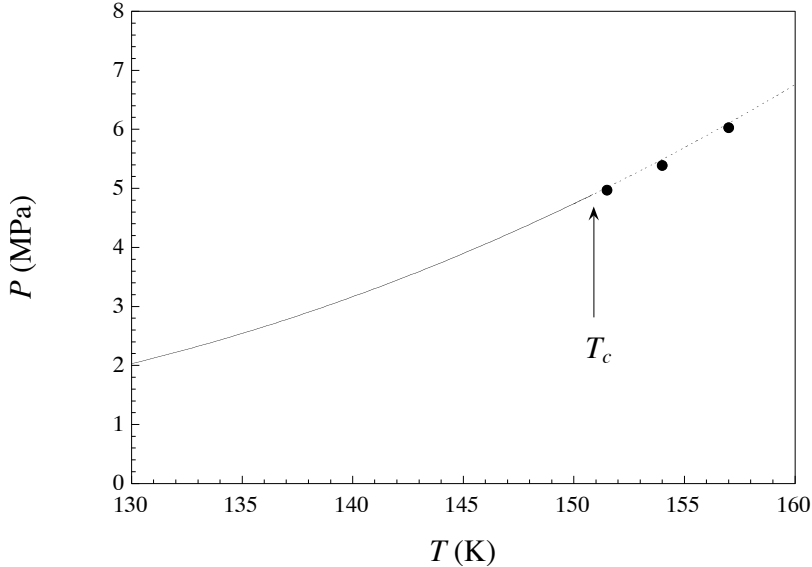}
\end{center}
\caption{Locus of points of minimum ion mobility. Solid line: coexistence curve \cite{rabinovich1988}.\label{fig:extrcoex}}
\end{figure}

A new analysis of the data 
reveals that the locus of the maximum mobility defect lies on the extrapolation of the coexistence curve into the single phase region, as shown in fig. \ref{fig:extrcoex}. The same occurs for $^{3}$He$^{+}$ ions in $^{3}$He near the critical point for $T_{c}\approx 3.3\,$K \cite{cantelli1968}.The gas appears reminiscent of the nearby coexistence curve even for $T>T_{c}.$ Droplets of superheated liquid are formed in the gas because of critical fluctuations. Electrostriction makes the local pressure around the ion  rise through the extrapolation of the coexistence line into the region of superheated liquid that the strong ion field stabilizes thermodynamically. This situation is also similar to the case of $^{4}$He$^{+}$ ions near the melting transition in which the $\lambda$-line is extrapolated into the region of supercooled, electrostrictively-stabilized liquid surrounding the ion \cite{scaramuzzi1977}. 

To get a microscopic insight of the ionic transport process, classical Molecular Dynamics (MD) simulations have been carried out in which the Newtonian equations of motion are numerically integrated for suitable interaction potentials by using freely available Fortran codes \cite{allen1992}. 
A system of 100 Ar atoms plus 1 O$_{2}^{-}$ ion has been simulated as a function of $N$ for $T=151.5\,$K. The Ar--Ar interaction potential is found in literature \cite{aziz1993}. The dominant long-range part of the O$_{2}^{-}-$Ar potential can be calculated by the knowledge of the polarizabilities of 
O$_{2}^{-}$ and Ar and of the quadrupole moment of Ar \cite{maitland}. The repulsive part of the O$_{2}^{-}-$Ar potential has been modeled as the interaction of the neutral species with an infinite wall endowed with a local charge density equal to the electronic charge density of the ion \cite{barker1983}. 

\begin{figure}[htbp]
\begin{center}
\includegraphics*[width=7.5cm,angle=90]{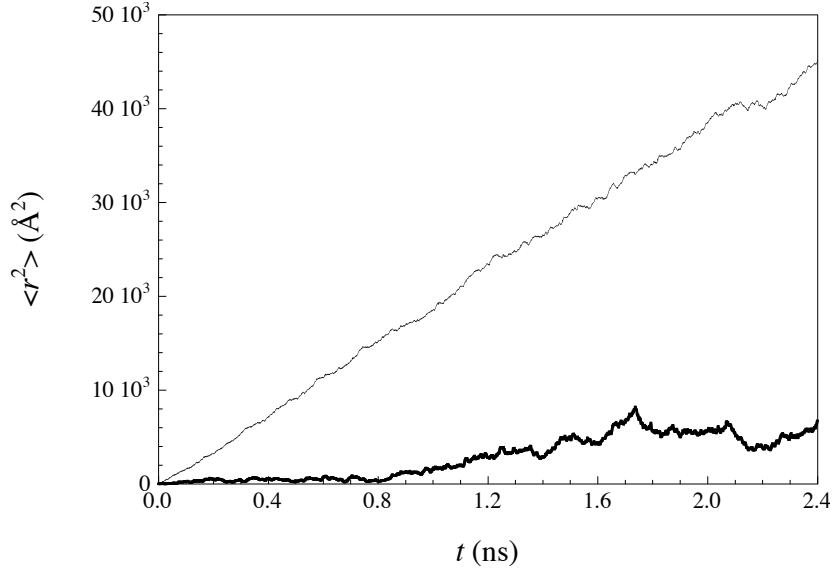}
\end{center}
\caption{$\langle r^{2}\rangle$ vs $t$ for $N=6.5\,$nm$^{-3}$ and $T=151.5\,$K.  Ar: thin line. O$_{2}^{-}:$ thick line.\label{fig:rms}}
\end{figure}

The equations of motion are integrated by using the Verlet algorithm \cite{allen1992} in time steps of $\tau=100 \tau_{i}=2.42 \cdot 10^{-15}\,$s, where $\tau_{i}= 4\pi \epsilon_{0}a_{0}\hbar/e^{2} $ is the timescale in atomic units (a.u.). After an initial equilibration time, the system evolves freely. The coordinates of all particles are recorded every time interval $t_{n}=(500 \tau) n, \, n=1,\ldots n_{f},$  spanning a total simulation time $\approx 2.4\,$ns for $n_{f}=2000$. 
In fig. \ref{fig:rms} we report a typical example of the mean square displacement $\langle r^{2}\rangle $ of the ion and of Ar for $N=6.5\,$nm$^{-3}$ and $T=151.5\,$K. For Ar, it is less noisy because of the higher statistical significance due to the averaging over 100 atoms. 
 As expected, the ion is less mobile than the neutral atoms because the enhancement of its interaction with atoms due to its electric field.
 
\begin{figure}[htbp]
\begin{center}
\includegraphics*[width=7.5cm,angle=90]{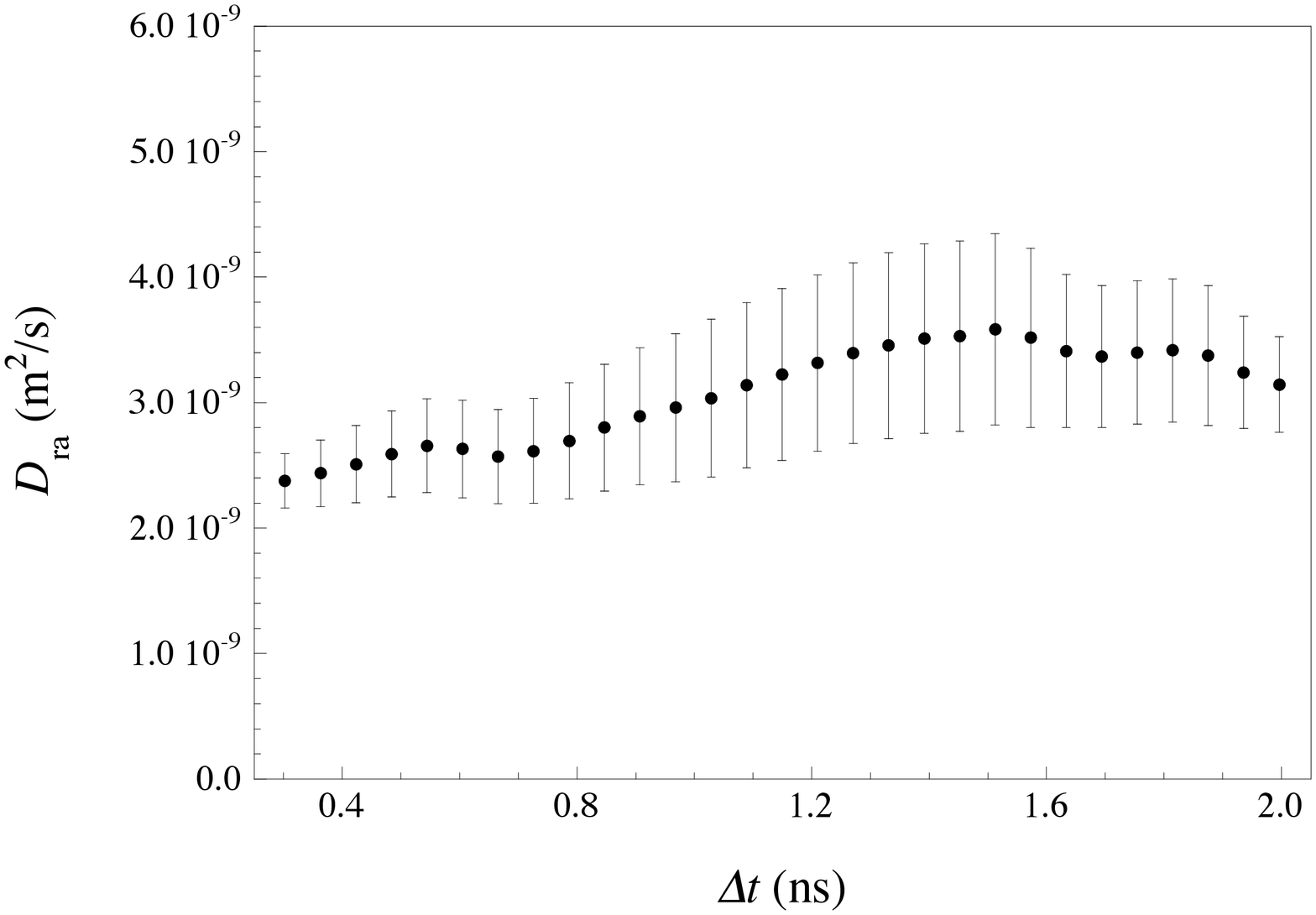}
\end{center}
\caption{$D_{\mathrm{ra}}$ vs $\mathit{\Delta}t.$
\label{fig:ddt}}
\end{figure}

 The diffusion coefficient $D$ is obtained as 
$D= \lim
_{\mathit{\Delta} t \to \infty}
  \langle r^{2}\left(\mathit{\Delta} t \right) \rangle/{6\mathit{\Delta} t }.
$ 
In order to improve the statistical significance of the data, a running average of the diffusion coefficient $D_{\mathrm{ra}}$ is first calculated by evaluating
$\langle r^{2}\rangle $ for each $\mathit{\Delta} t $ by moving the time origin across the whole simulation and averaging over the results before averaging over the sampling window $\mathit{\Delta} t .$
In fig. \ref{fig:ddt} a typical result for the running average of $D_{\mathrm{ra}}$ is plotted as a function of the sampling window $\mathit{\Delta} t. $ The error bars are the results of averaging over the different time origins. For practical purposes, the limit in the definition of $D$ is replaced by the mean of the running average $D_{\mathrm{ra}}.$
Finally, the Nernst--Einstein relationship $D/\mu_{i}=k_{\mathrm{B}}T/e$ is used to calculate the ion mobility.


\begin{figure}[htbp]
\begin{center}
\includegraphics*[width=7.5cm,angle=-90]{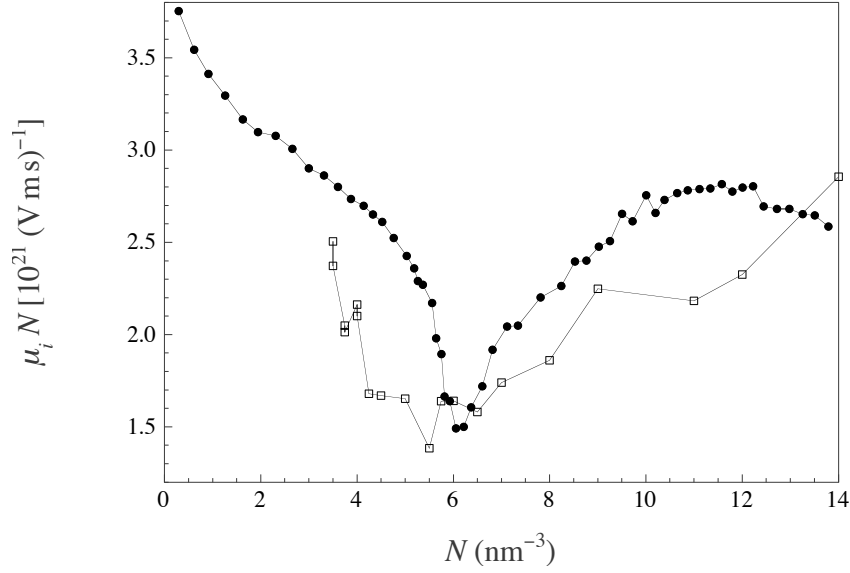}
\end{center}
\caption{Comparison between experiment (closed points) and MD simulations (open points). The error bars are not shown for graphical clarity. Solid lines: eyeguides.\label{fig:cfrdatsim}}
\end{figure}

In fig. \ref{fig:cfrdatsim} the MD results for $T=151.5\,$K are compared with the experimental data. The quantitative agreement between data and simulation is very good. The simulated $\mu_{i}$ shows the defect exactly at the same density of the experiment. The density width of the mobility defect calculated in the simulations is larger than the experimental one. It is believed that this effect is due to the fact that the temperature during simulation is allowed to substantially fluctuate in order for energy to be constant.

\begin{figure}[htbp]
\begin{center}
\includegraphics*[width=7.5cm,angle=90]{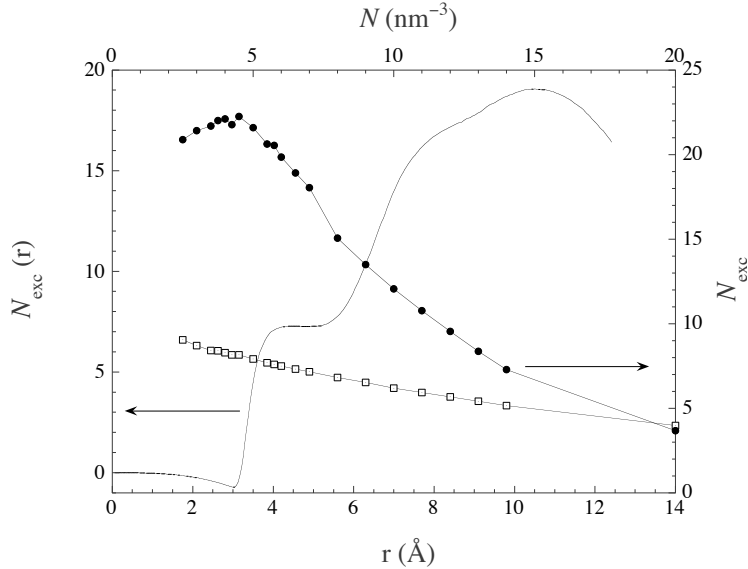}
\end{center}
\caption{$N_{\mathrm{exc}}$ vs $r$ for $N=6.5\,$nm$^{-3}$ and $T=151.5\,$K (left- and bottom axes). $N_{\mathrm{exc}}$ vs $N$ for $T=151.5\,$K. Squares: 1st solvation shell. Circles: 2nd solvation shell (top- and right axes).
\label{fig:nexcr}}
\end{figure}

Disappointingly, however, MD simulations do not enlighten the physical reasons of their success. Actually, MD calculations also provide the ion--atom pair distribution function $g(r)$ that gives important information on the structure of the ion environment. In particular, the excess number $N_{\mathrm{exc}}$ of atoms surrounding the ion can be calculated. In fig. \ref{fig:nexcr} $N_{\mathrm{exc}}$ is reported as a function of the distance $r$ from the ion for $N$ close to that of the mobility minimum. The formation of the first solvation shell is clearly indicated by the plateau starting at $r\approx 4\,$\AA, while a diffuse second solvation shell is observed for $r\approx 9\,$\AA.
In fig.~\ref{fig:nexcr}, the number of excess atoms contained in the 1st and 2nd solvation shells is also shown as a function of $N.$ Interestingly enough, $N_{\mathrm{exc}}$ in the 2nd solvation shell is strongly peaked for $N< N_{c},$ thus suggesting that electrostriction  efficiently modifies the ion environment when the unperturbed gas density is well below $N_{c}.$ On the other hand, nothing anomalous shows up around the density of the maximum mobility defect. We believe that carrying out more MD simulations will improve their statistical significance and thus shed light on the physical picture beneath the experiment.


\end{document}